%% file: ubracket_anon.tex
\documentclass[twoside,
12pt,a4paper]{article}
\usepackage[marked,extnum,printedin]{myart2k}
\input{mydef}

\ppnum{}{LEEDS-PURE-MATH-2000-23\\
\href{http://arXiv.org/abs/math-ph/0007030}%
{arXiv:math-ph/0007030}}%
{2000}

\newcommand{\anti}{\mathcal{A}}
\newcommand{\ub}[3][]{\left\{\!#1\left[#2,#3\right]\!#1\right\}}
\newcommand{\myhbar}{\hbar}

\title{Quantum and Classic Brackets}
\author[Vladimir V. Kisil]%
{\href{http://amsta.leeds.ac.uk/~kisilv/}{Vladimir V. Kisil}}

\address{%
School of Mathematics\\
University of Leeds\\
Leeds LS2\,9JT\\
UK}

\email{\href{mailto:kisilv@amsta.leeds.ac.uk}{kisilv@amsta.leeds.ac.uk}}

\urladdr{\href{http://amsta.leeds.ac.uk/~kisilv/}%
{http://amsta.leeds.ac.uk/\~{}kisilv/}}

\date{July 25, 2000}
\begin{document}

\maketitle

\begin{abstract}
  We describe an $p$-\-mechanical~\cite{Kisil96a,Kisil94e,Prezhdo-Kisil97}
  brackets which generate quantum
  (commutator) and classic (Poisson) brackets in corresponding
  representations of the Heisenberg group.  We
  \emph{do not} use any kind of semiclassic approximation or limiting
  procedures for $\hbar \rightarrow 0$.
\end{abstract}
  \AMSMSC{81R05}{81P05, 22E70, 43A65}

{\small\tableofcontents}

\section{Introduction}
\label{sec:introduction}

The purpose of this short announcement is to describe a ``brackets''
in $p$-mechanical setting~\cite{Kisil96a,Kisil94e,Prezhdo-Kisil97}
which generates both classic (Poisson) and 
quantum (commutator) brackets. Consequently we are able to derive
dynamical equation in classic and quantum cases from the same
consistent source. 

The principal step in transition from Lagrangian to Hamiltonian
mechanics is introduction by means of the Legendre transform
\emph{new independent 
  variables}---coordinates and momentums---instead of coordinates and
depending from them their time derivatives---velocities $\dot{q}$.
Similarly $p$-mechanical
construction~\cite{Kisil96a,Kisil94e,Prezhdo-Kisil97} is based on
introduction by means of the Fourier transform new variables $(s,x,y)$
such that $(x,y)$ is Fourier dual to $(q,p)$ and $s$ is Fourier dual
to the Planck constant $\hbar$. It appeared that points $(s,x,y)$ are
elements of the Heisenberg group
$\Space{H}{n}$~\cite{Howe80a,Howe80b,MTaylor86} (see
also~\eqref{eq:H-n-group-law}).

It is known since works of von Neumann that the Heisenberg picture of
quantum mechanics is generated by infinite dimensional non-commutative
irreducible unitary representations of
$\Space{H}{n}$~\cite{Howe80b}. But one-dimensional (commutative!)
unitary representations of $\Space{H}{n}$ are oftenly unemployed. It is
shown within $p$-mechanical framework that these one-dimensional
representations contain classic dynamics exactly in the same way as
infinite-dimensional ones---quantum. 

An important feature of our approach that we
\emph{do not} use any kind of semiclassic approximation or limiting
procedures for $\hbar \rightarrow 0$, the classic picture is not any
more an imperfect shade of a quantum description. 

Here we present a
$p$-mechanical version of brackets and a dynamical equation generated
by them. Our considerations is illustrated by a simple example of
harmonic oscillator. More involved examples allowing mix quantum and
classic components within one system will be presented elsewhere. 
 
\input{ubracket0a.tex}

{\small
\bibliographystyle{klunum}
\bibliography{abbrevmr,akisil,analyse,aphysics}
}

\end{document}

%% file: ubracket0a.tex

\section{Preliminaries}

\subsection{Groups and Their Representations}
\label{sec:groups-their-repr}

We consider  $\FSpace{L}{2}(\Space{R}{n})$ equipped with the scalar
product 
\begin{eqnordisp}[eq:L2-product]
  \scalar{f}{g} = \frac{1}{\pi^{n/2}} \int_{\Space{R}{n}} f(x)
  \bar{g}(x)\,dx. 
\end{eqnordisp}
Through the paper we use the standard notation for the Fourier
transform:
\begin{eqnordisp}{}
  [\fourier{}  f](\myhbar)= \hat{f}{\myhbar}
  \sqrt{2\pi}\int\limits_{-\infty}^\infty f(s)\,  e^{-i s\myhbar}\,ds
\end{eqnordisp}

Let $G$ be a group with an invariant measure $dg$.
$\FSpace{L}{1}(G,dg)$  could be upgraded from a
linear space to an algebra with the convolution
multiplication:
\begin{eqnordisp}[eq:de-convolution]
  (k_1 * k_2) (g) = \int_G k_1(h)\, k_2(h^{-1}g)\,dh =
  \int_G k_1(gh^{-1})\, k_2(h)\,dh
\end{eqnordisp}

Let $\rho$ be a representation of $G$~\cite[Chap.~1]{MTaylor86},
we will work mainly with unitary irreducible ones. We could extend
$\rho$ to $\FSpace{L}{1}(G,dg)$ by the formula:
\begin{eqnordisp}[eq:L1-repr]
  \rho(k)= \int_G k(g)\rho(g)\,dg.
\end{eqnordisp}
From the general properties of representations of Lie
groups~\cite[Chap.~1, (2.17)]{MTaylor86} we have:
\begin{eqnordisp}[eq:repr-basic]
  \rho(k_1)+\lambda\rho(k_2)= \rho(k_1+\lambda k_2),\qquad
  \rho(k_1)\,\rho(k_2)= \rho(k_1*k_2).
\end{eqnordisp}
This could be reinforced in the following statement. 
\begin{lem}[Algebraic Inheritance]
  \label{le:inheritance}
  Let $p(a_1,a_2,\ldots,a_n)$ be a polynomial in non-commuting
  arguments $a_1$, $a_2$,\ldots, $a_n$. Let functions $k_1$, $k_2$,
  \ldots, $k_n$ from $\FSpace{L}{1}(G)$ satisfy to the identity
  \begin{eqnordisp}
    p(k_1,k_2,\ldots,k_n)=0,
  \end{eqnordisp}
  where multiplication is defined as the group convolution on $G$. Then
  \begin{eqnordisp}
    p(\rho(k_1),\rho(k_2),\ldots,\rho(k_n))=0
  \end{eqnordisp}
  for an arbitrary representation $\rho$ of $G$.
\end{lem}

\subsection[Heisenberg Group and Representations]{The 
Heisenberg Group $\Space{H}{n}$ and Its Representations}
\label{sec:math-found}

Let $(s,x,y)$, where $x$, $y\in \Space{R}{n}$ and $s\in\Space{R}{}$, be
an element of the Heisenberg group
$\Space{H}{n}$~\cite{Howe80a,Howe80b,MTaylor86}. The group law on
$\Space{H}{n}$ is given as follows:
\begin{eqnordisp}[eq:H-n-group-law]
  (s,x,y)*(s',x',y')=(s+s'+\frac{1}{2}(xy'-x'y),x+x',y+y').
\end{eqnordisp} 

For our purpose we need all irreducible representations of the group
\Heisen{n}. They are given by the following famous theorem:
\begin{thm}[Stone-von Neumann]
  \label{th:Stone-von-Neumann}
  \cite[\S~18.4]{Kirillov76}, \cite[\S~1.2]{MTaylor86} All unitary
  irreducible representations of the Heisenberg group $\Space{H}{n}$
  up to unitary equivalence are as follows
  \begin{enumerate}
  \item For any $\myhbar \in (0,\infty)$ the Schr\"odinger irreducible
    noncommutative unitary representations in
    $\FSpace{L}{2}(\Space{R}{n})$
    \begin{equation}\label{eq:stone1}
      \rho_{\pm\myhbar}(s,x,y)=e^{i(\pm s\cdot\myhbar I
        \pm x\cdot \myhbar^{1/2}M +y\cdot\myhbar^{1/2}D)},
    \end{equation}
    where $xM$ and $yD$ are such unbounded self-adjoint operators on
    $\FSpace{L}{2}(\Space{R}{n})$:
    \begin{eqnarray}
      \label{eq:M-definition}
      (x \cdot \myhbar^{1/2}M)u(v)&=&\myhbar^{1/2}\sum
      x_{j}v_{j}u(v),\\
      \label{eq:D-definition}
      (y \cdot \myhbar^{1/2}D)u(v)&=&\frac{\myhbar^{1/2}}{i} 
      \sum y_{j}\frac{\partial u}{\partial v_{j}}.
    \end{eqnarray}
    Representation \eqref{eq:stone1} acts on a function $u(v)$ as
    follows: 
    \begin{eqnordisp}[eq:stone1-act]
      \rho_{\pm\myhbar}(s,x,y) u(v)= e^{i(\pm (s+xy/2)\cdot\myhbar I
        \pm x\cdot \myhbar^{1/2}v)} u(v+\myhbar^{1/2}y)
    \end{eqnordisp}
  \item For $(q,p)\in\Space{R}{2n}$ commutative
    one-dimensional representations on \Space{C}{}:
    \begin{equation}
      \label{eq:stone2}
      \rho_{(q,p)}(s,x,y) u=e^{i(q x+p y)}u,\ u\in\Space{C}{}.
    \end{equation}
  \end{enumerate}
\end{thm} 

In some sense~\cite{Kisil96a} the last
representations~\eqref{eq:stone2} correspond to the case
$\myhbar=0$. While other representations of $\Space{H}{n}$ could be
transformed to the above ones by unitary operators it is better
sometime to stay with alternative forms tailored to particular
models. For example, the Segal-Bargmann
representation~\cite{Segal60,Bargmann61} is well suited for quantum
field theory and its relation to the Schr\"odinger
representation~\eqref{eq:stone1} illuminate many results in analysis
and quantum theory~\cite{Howe80b}.

Representations
(\ref{eq:stone1}--\ref{eq:stone2}) generate accordingly to
\eqref{eq:L1-repr} representations of convolution algebra
$\FSpace{L}{1}(\Space{H}{n})$ 
expressed by formulas \cite[Chap.~1, (3.9)]{MTaylor86}:
\begin{eqnarray}
  \rho_{\pm\myhbar}[k(s,x,y)]&=&\hat{k}(\pm\myhbar,
  \pm\myhbar^{1/2}M,\myhbar^{1/2}D),\label{eq:cstone1}\\
  \rho_{(q,p)}[k(s,x,y)] &=&\hat{k}(0,q,p)\label{eq:cstone2}.
\end{eqnarray} 
The right side of~\eqref{eq:cstone1} specifies a
pseudo-differential operator (PDO)~\cite{Hormander85,Shubin87} with
the Weyl symbol $\hat{k}(\pm\myhbar,
\pm\myhbar^{1/2}x,\myhbar^{1/2}\xi)$. Such a PDO with a symbol 
$a(v,\nu)$ defined by:
\begin{equation}
  a_{\tau}(M,D)\, u(v) =(2\pi)^{-N} \int_{\Space{R}{N}}
  \int_{\Space{R}{N}}
  e^{i<v-u,\nu>}\, a(\tau u+(1-\tau)v,\nu)\, u(u)\, d\nu\,du.
  \label{eq:Weyl-PDO}
\end{equation}
The right side of~\eqref{eq:cstone2} is just a constant from
\Space{C}{}.

Using~\eqref{eq:repr-basic} with $\rho$ equal either to
$\rho_\myhbar$~\eqref{eq:stone1} or to
$\rho_{(q,p)}$~\eqref{eq:stone2} we obtain
\begin{eqnordisp}[eq:repr-commutant]
  \rho(k_1*k_2-k_2*k_1)=
  \left\{
    \begin{array}{ll}
      [K_1,K_2]=K_1K_2-K_2K_1, & \quad \rho=\rho_\myhbar,\ \myhbar\neq 0;\\
      0, & \quad \rho=\rho_{(q,p)},
    \end{array}
  \right.
\end{eqnordisp}
where operators $K_1$ and $K_2$ are Weyl PDO defined
by~\eqref{eq:cstone1} for functions $k_1$ and $k_2$ respectively. 

\section{Quantum and Classic Brackets}
\label{sec:quant-class-mech}

\subsection[p-mechanical Brackets]{$p$-\-mechanical Brackets 
and Its Quantum and Classic Representations}
\label{sec:univ-brackets} 

Let $\FSpace[v]{L}{1}(\Space{R}{})$
be the linear subspace of $\FSpace{L}{1}$  functions on
$\Space{R}{}$ such that:
\begin{eqnordisp}
  \lim_{s\rightarrow -\infty} s\int\limits_{-\infty}^s f(t)\,dt  =  0,
  \quad \textrm{ and } \quad
  \lim_{s\rightarrow \infty} s\int\limits^{\infty}_s f(t)\,dt  =  0.
\end{eqnordisp}
A non-trivial function from $\FSpace[v]{L}{1}(\Space{R}{})$ is, for
example, $xe^{-x^2}$.
The following could be easily seen (cf.~\cite[\S~IV.1.1 and
\S~IV.2.3]{KirGvi82}). 
\begin{lem}
  \begin{enumerate}
  \item \label{it:lo-closed}
    $\FSpace[v]{L}{1}(\Space{R}{})$ is a \emph{closed ideal} in
    convolution algebra $\FSpace{L}{1}(\Space{R}{})$.
  \item \label{it:lo-fourier}
    The Fourier transform of functions from
    $\FSpace[v]{L}{1}(\Space{R}{})$ are among continuous
    functions such that $\hat{f}(0)=0$.
\end{enumerate}
\end{lem}
Let $\anti$ be an
anti-derivation---linear unbounded operator from
$\FSpace[v]{L}{1}(\Space{R}{})$ onto the space of integrable
functions on $\Space{R}{}$ defined by the formula:
\begin{eqnordisp}[eq:anti-defn]
  [\anti f](s) = \int\limits_{-\infty}^s f(t)\,dt 
  = \int\limits_{-\infty}^\infty \chi(s-t)\,f(t)\,dt,
\end{eqnordisp}
where $\chi(t)$ is the Heaviside function:
\begin{eqnordisp}[eq:Heaviside]
  \chi(t)= \left\{ 
    \begin{array}{ll}
      0, & \textrm{ if } t\leq 0;\\
      1, & \textrm{ if } t> 0.\\
    \end{array}
  \right.
\end{eqnordisp}
From the definition it follows that:
\begin{lem}
  The antiderivative $\anti$ enjoys the following properties:
  \begin{enumerate}
  \item \label{it:A-null}
    $\anti \mathbf{0} = \mathbf{0}$, where $\mathbf{0}$ is the
    function identically equal to $0$. The function $\mathbf{0}$ is
    the only element of the kernel of $\anti$: $\ker
    \anti=\{\mathbf{0}\}$; 
  \item \label{it:A-commute}
    $\anti$ commutes with all shifts $f(s)\rightarrow f(s+a)$ and
    their linear combi\-na\-tions---convolution operators on
    $\Space{R}{}$. 
  \item \label{it:A-limits}
    For $ f\in\FSpace[v]{L}{1}(\Space{R}{})$ the limits at infinity
    vanish:
    \begin{eqnordisp}[eq:anti-limits]
      \lim_{s\rightarrow-\infty} [\anti f](s) =
      \lim_{s\rightarrow\infty} [\anti f](s) = 
      \lim_{s\rightarrow-\infty} s[\anti f](s) =
      \lim_{s\rightarrow\infty} s[\anti f](s) = 0. 
    \end{eqnordisp}
  \item \label{it:A-closed}
    If $f_1$, $f_2\in \FSpace[v]{L}{1}(\Space{R}{})$ then 
    $\anti (f_1*f_2)=(\anti f_1)*f_2=f_1*(\anti f_2)$ is again in 
    $\FSpace[v]{L}{1}(\Space{R}{})$.
  \end{enumerate}
\end{lem}
From integration by parts:
\begin{eqnordisp}
  \int\limits_{-\infty}^\infty [\anti f](s)\, e^{-i s \myhbar}\, ds =
  \left.
    [\anti f](s)\,\frac{e^{-i s \myhbar}}{-i\myhbar}
  \right|_{-\infty}^\infty 
  -\int\limits_{-\infty}^\infty f(s)\, \frac{e^{-i s \myhbar}}{-i\myhbar}\, ds
\end{eqnordisp}
and~\eqref{eq:anti-limits} we obtain:
\begin{eqnordisp}[eq:fourier-anti]
  \fourier{} [\anti f](\myhbar)=
  \left\{
    \begin{array}{ll}
      \frac{1}{i\myhbar}[\fourier{} f](\myhbar), & 
      \myhbar \neq 0;\vspace{2mm}\\
      -\sqrt{2\pi}\int\limits_{-\infty}^\infty f(s)\, s\,ds, & \myhbar = 0,
    \end{array}
  \right.
\end{eqnordisp}
for $f(s)\in\FSpace[v]{L}{1}(\Space{R}{})$. In fact we could take the
last formulae as a definition of the operator $\anti$.
\begin{defn}
  \label{de:u-brackets}
  The \emph{$p$-\-mechanical brackets} of two functions $k_1(s,x,y)$,
  $k_2(s,x,y)$ on the Heisenberg $\Space{H}{n}$ are defined as
  follows: 
  \begin{eqnordisp}[eq:u-brackets]
    \ub{k_1}{k_2}= \anti(k_1*k_2-k_2*k_1),
  \end{eqnordisp}
  where $*$ denotes the group convolution on $\Space{H}{n}$ of two
  functions and $\anti$ acts as antiderivative with respect of the
  variable $s$. 
\end{defn}

This definition of the $p$-\-mechanical bracket has sense if
$k_{1,2}(s,x_0,y_0) \in \FSpace[v]{L}{1}(\Space{R}{})$ for
any fixed $x_0, y_0 \in \Space{R}{n}$. 
Due to Lemma~\ref{it:A-closed}
the $p$-brackets of two such functions is again in
$\FSpace[v]{L}{1}(\Space{R}{})$, thus $\anti$ is meaningful
in~\eqref{eq:u-brackets}. While this completely serves the purpose of
the present paper future extensions of the
Definition~\ref{de:u-brackets} are also possible. Note also, that we
put $\FSpace[v]{L}{1}$-condition only with respect to variable $s$;
variables $x$ and $y$, which are Fourier-dual to physical coordinates
and momentum, are unrestricted.
\begin{lem}
  \label{le:ub-Lie}
  The $p$-\-mechanical brackets~\eqref{eq:u-brackets} have the following 
  properties
  \begin{enumerate}
  \item They are linear.
  \item They are antisymmetric $\ub{k_1}{k_2}=-\ub{k_2}{k_1}$.
  \item They satisfy to the Jacoby identity 
    \begin{eqnordisp}[eq:Jacoby]
      \ub{\ub{k_1}{k_2}}{k_3}+\ub{\ub{k_2}{k_3}}{k_1}
      +\ub{\ub{k_3}{k_1}}{k_2}=0.
    \end{eqnordisp}
  \item They are a derivation, i.e. satisfy to the Leibniz rule:
    \begin{eqnordisp}[eq:Leibniz]
      \ub{k_1*k_2}{k_3}=\ub{k_1}{k_3}*k_2+k_1*\ub{k_2}{k_3}.
    \end{eqnordisp}
  \end{enumerate}
\end{lem}
\begin{proof}
  The linearity and antisymmetric properties are obvious. Two other
  properties are secured because 
  \begin{enumerate}
  \item $\anti$ commutes with convolutions (Lemma~\ref{it:A-commute}) and
    sends zero function to itself (Lemma~\ref{it:A-null});
  \item The commutator $k_1*k_2-k_2*k_1$ satisfies both to Jacoby and
        Leibniz identity.
  \end{enumerate}
  For example the Leibniz identity could be verified as follows:
  \begin{eqnarray}
    \lefteqn{\ub{k_1*k_2}{k_3}  =  \anti(k_1*k_2*k_3-k_3*k_1*k_2)} \nonumber\\
    & = & \anti(k_1*k_2*k_3-k_1*k_3*k_2+k_1*k_3*k_2-k_3*k_1*k_2)
    \nonumber\\
    & = & \anti(k_1*k_2*k_3-k_1*k_3*k_2)
    +\anti(k_1*k_3*k_2-k_3*k_1*k_2) \qquad\label{eq:Leibniz-1}\\
    & = & k_1*\anti(k_2*k_3-k_3*k_2)
    +\anti(k_1*k_3-k_3*k_1)*k_2 \label{eq:Leibniz-2}\\
    & = & k_1*\ub{k_2}{k_3}+\ub{k_1}{k_3}*k_2, \nonumber
  \end{eqnarray}
  where \eqref{eq:Leibniz-1} follows from the linearity of $\anti$ and
  \eqref{eq:Leibniz-2} is a consequence of Lemma~\ref{it:A-commute}.
\end{proof}

Now we describe image of the brackets under representations of
$\Space{H}{n}$. 
\begin{prop}
  \label{pr:bracket-repr}
  The images of $p$-\-mechanical brackets~\eqref{eq:u-brackets} under
  infinite dimensional representations $\rho_\myhbar$, $\myhbar \neq 0$ and
  finite dimensional representations $\rho_{(q,p)}$ are quantum
  commutant and Poisson brackets of functions $\hat{k}_1$ and
  $\hat{k}_2$ respectively:
  \begin{eqnordisp}[eq:u-bra-repr]
    \rho(\ub{k_1}{k_2})=
    \left\{
    \begin{array}{ll}
      \displaystyle \frac{1}{ i\myhbar}[\hat{k}_1,\hat{k}_2]=
      \frac{1}{ i\myhbar}(K_1K_2-K_2K_1), & \ 
      \rho=\rho_\myhbar,\ \myhbar\neq 0;\vspace{2mm}\\
      \displaystyle \{\hat{k}_1,\hat{k}_2\}= 
      \frac{\partial \hat{k}_1}{\partial q}
      \frac{\partial \hat{k}_2}{\partial p}
      - \frac{\partial \hat{k}_1}{\partial p}
      \frac{\partial \hat{k}_2}{\partial q}, & \ 
      \rho=\rho_{(q,p)}.
    \end{array}
    \right.
  \end{eqnordisp}
\end{prop}
\begin{proof}
  The proof is a straightforward calculation
  using~\eqref{eq:fourier-anti}. We will carry them separately for
  cases of $\myhbar\neq 0$ and $\myhbar=0$. 

  Let $\rho=\rho_\myhbar$, $\myhbar\neq 0$. Then:
  \begin{eqnarray}
    \lefteqn{\rho_\myhbar(\ub{k_1}{k_2})  =  
      \int_{\Space{H}{n}} \ub{k_1}{k_2}\!(g)\, \rho_\myhbar(g)\,dg }
    \nonumber\\
    & = & \int_{\Space{H}{n}} \anti(k_1*k_2-k_2*k_1)(s,x,y)\, 
    e^{i(\pm\myhbar sI\pm\myhbar^{1/2} xM+\myhbar^{1/2}yD)}\,dg 
    \nonumber\\  
    & = &  \frac{1}{i\myhbar}
    \int_{\Space{H}{n}} (k_1*k_2-k_2*k_1)(s,x,y)\,
    e^{i(\pm\myhbar s I\pm\myhbar^{1/2}
      xM+\myhbar^{1/2}yD)}
    \,dg 
    \label{eq:h-trans-1}\\  
    & = &  \frac{1}{i\myhbar}[K_1,K_2],  \label{eq:h-trans-3}
  \end{eqnarray}
  where the line~\eqref{eq:h-trans-1} follows from the first case
  in~\eqref{eq:fourier-anti} and~\eqref{eq:h-trans-3} is exactly the
  first case in~\eqref{eq:repr-commutant}.

  The second case $\rho=\rho_{(q,p)}$ (symbolically corresponding
  to``$\myhbar=0$'') is also not difficult but somehow
  longer: 
  \begin{eqnarray}
    \lefteqn{ 
      \rho_{(q,p)}(\ub{k_1}{k_2}) 
      = \int_{\Space{H}{n}} \ub{k_1}{k_2}\!(g)\, \rho_{(q,p)}(g)\,dg }
    \nonumber\\
    \qquad
    & = & \int_{\Space{H}{n}} \anti(k_1*k_2-k_2*k_1)(s,x,y)\, 
    e^{i(qx+py)}\,dg 
    \nonumber\\
    & = & \int_{\Space{H}{n}} (k_2*k_1-k_1*k_2)(s,x,y)\, 
    s e^{i(qx+py)}\,dg 
    \label{eq:pq-trans-1}\\
    & = & \int_{\Space{H}{n}} \int_{\Space{H}{n}}\! \left(
    k_2(s',x',y')\,k_1(s-s'+\frac{x'y-xy'}{2},x-x',y-y') \right.
    \nonumber \\ 
    && \qquad \left. -k_1(s',x',y')\,k_2(s-s'+\frac{x'y-xy'}{2},x-x',y-y')
    \right)dg'
    \nonumber \\ 
    && \quad \times s e^{i(qx+py)}\,dg \nonumber
  \end{eqnarray} 
  We use the second case of~\eqref{eq:fourier-anti}
  to obtain~\eqref{eq:pq-trans-1}. Now let us change variables 
  \begin{eqnordisp}[eq:change-var]
    \begin{array}{rcl}
      x''=x-x',& y''=y-y', & s''=s-s'+\frac{x'y-xy'}{2}; \vspace{2mm}\\
      x =x''+x',& y=y''+y' & s=s''+s'+\frac{x''y'-x'y''}{2},
    \end{array}
  \end{eqnordisp}
  and continue   the above calculations:
  \begin{eqnarray}
    & = & \int_{\Space{H}{n}} \int_{\Space{H}{n}}\left( 
    k_2(s',x',y')\,k_1(s'',x'',y'')  
    -k_1(s',x',y')\,k_2(s'',x'',y'') \right)
    \nonumber \\ 
    && \quad  \times
    \left(s''+s'+\frac{x''y'-x'y''}{2}\right)
    e^{i(q(x''+x')+p(y''+y'))}
    \,dg'\,dg'' \nonumber \\
    & = & \int_{\Space{H}{n}} \int_{\Space{H}{n}}\left( 
    k_2(s',x',y')\,k_1(s'',x'',y'')
    -k_1(s',x',y')\,    k_2(s'',x'',y'') \right)
    \label{eq:pq-trans4} \\ 
    && \quad  \times
    \left(s''+s'\right)e^{i(qx'+py')} e^{i(qx''+py'')}
    \,dg'\,dg'' \label{eq:pq-trans5}\\
    &&{}+\int_{\Space{H}{n}}\int_{\Space{H}{n}}\!\!\left( 
    k_2(s',x',y')\,k_1(s'',x'',y'')
    -k_1(s',x',y')\,    k_2(s'',x'',y'') \right)\qquad\quad
    \label{eq:pq-trans2} \\ 
    && \quad  \times
    \frac{x''y'-x'y''}{2}\,e^{i(qx'+py')} e^{i(qx''+py'')}
    \,dg'\,dg'' \label{eq:pq-trans3}
  \end{eqnarray} 
  Interchanging primed and double primed variables
  in~\eqref{eq:pq-trans4}--\eqref{eq:pq-trans5} we conclude that the
  integral is equal to itself with the opposite sign and thus
  vanish. In contrast such an interchange in the
  integral~\eqref{eq:pq-trans2}--\eqref{eq:pq-trans3} lead to 
  a continuation of~\eqref{eq:pq-trans4}--\eqref{eq:pq-trans3}:
  \begin{eqnarray}
    & = & \int_{\Space{H}{n}} \int_{\Space{H}{n}}
    \left(k_2(s',x',y')\,k_1(s'',x'',y'')
      -k_1(s',x',y')\,k_2(s'',x'',y'')\right)   
    \nonumber\\
    && \quad  \times
    x''y'\,e^{i(qx'+py')} e^{i(qx''+py'')}
    \,dg'\,dg''\nonumber\\
    & = & \int_{\Space{H}{n}} k_2(s',x',y')\,y'\,e^{i(qx'+py')}
    \,dg' 
    \int_{\Space{H}{n}} k_1(s'',x'',y'')\,x'' e^{i(qx''+py'')}
    \,dg'' \nonumber\\
      &&{}-\int_{\Space{H}{n}} k_1(s',x',y')\, y'e^{i(qx'+py')}
      \,dg'
      \int_{\Space{H}{n}}k_2(s'',x'',y'')\,x'' e^{i(qx''+py'')}
    \,dg''\nonumber\\
    &=&\frac{\partial \hat{k}_2(0,q,p)}{\partial p}\,
    \frac{\partial \hat{k}_1(0,q,p)}{\partial q}
    -\frac{\partial \hat{k}_1(0,q,p)}{\partial p}\,
    \frac{\partial \hat{k}_2(0,q,p)}{\partial q}
    \nonumber\\
    & = & \{ k_1, k_2\}.\nonumber
  \end{eqnarray}
  This finishes the proof.
\end{proof} 
\begin{rem}
  Let $S$, $X_j$, $Y_j$ $j=1,\ldots,n$ be vectors spanning the Lie algebra of
  $\Space{H}{n}$, i.e. $[X_j,Y_j]=S$ and all other commutators
  vanish. Consequently the only nontrivial
  $p$-brackets among those vectors are
  $\ub{X_j}{Y_k}=\delta_{jk}I$. By the algebraic inheritance
  (Lemma~\ref{le:inheritance}) we find the only non-trivial quantum
  and classic brackets:
  \begin{eqnordisp}{}
    \frac{1}{i\myhbar}[\rho_\myhbar(X_j),\rho_\myhbar(Y_k)]= I, \qquad
    \{\rho_{(q,p)}(X_j),\rho_{(q,p)}(Y_k)\} = I.
  \end{eqnordisp}
\end{rem}

The role of the antiderivative $\anti$ in~\eqref{eq:u-brackets} is
highlighted by a comparison of~\eqref{eq:repr-commutant}
and~\eqref{eq:u-bra-repr}. $\anti$ does not only insert the multiplier
$\frac{1}{i\myhbar}$ in quantum commutant, it also (and this is
essentially new in our construction) produces a
\emph{non-trivial classical representation} of the $p$-\-mechanical brackets.

The following corollary is very well known but we would like to
incorporate it in our scheme.
\begin{cor}
  \label{co:CP-properies}
  The quantum commutator and the Poisson brackets are linear,
  antisymmetric, and satisfy to the Jacoby~\eqref{eq:Jacoby} and
  Leibniz~\eqref{eq:Leibniz} identities. 
\end{cor}
\begin{proof}
  The properties follows from the corresponding properties of
  $p$-\-mechanical brackets (Lemma~\ref{le:ub-Lie}) and conservation of
  algebraic identities by representations
  (Lemma~\ref{le:inheritance}).{}  
\end{proof}

As a direct consequence of the Proposition~\ref{pr:bracket-repr} we
obtain the following statement.
\begin{thm}
  Let a function $f(t;s,x,y)$ defined on
  $\Space{R}{}\times\Space{H}{n}$ be a solution of the $p$-\-mechanical
  equation:
  \begin{eqnordisp}[eq:universal]
    \frac{d}{dt} f(t;s,x,y)= \ub{f}{H}
  \end{eqnordisp}
  with a ``Hamiltonian'' $H(s,x,y)$ on $\Space{H}{n}$. Then 
  \begin{enumerate}
  \item The operator $f_\myhbar(t;M,D)=[\rho_\myhbar f](t;M,D)$
    representing $f(t;s,x,y)$ under
    $\rho_\myhbar$~\eqref{eq:cstone1} is a solution of the Heisenberg 
    equation
    \begin{eqnordisp}[eq:Heisenberg]
      \frac{d}{dt} f_\myhbar(t;X,D)=
      \frac{1}{i \myhbar} [f_\myhbar,H_\myhbar],
    \end{eqnordisp} 
    with the Hamiltonian operator
    $H_\myhbar(M,D)=[\rho_\myhbar H](M,D)$ from~\eqref{eq:cstone1}.
  \item The function $f_0(t;q,p)=[\rho_{(q,p)} f]$ constructed
    by~\eqref{eq:cstone2} is a solution of the Hamilton equation:
    \begin{eqnordisp}[eq:Hamilton]
      \frac{d}{dt} f_0(t;q,p)= \{ f_0,H_0\},
    \end{eqnordisp} 
    where the Hamiltonian function $H_0(q,p)=[\rho_{(q,p)} H]$ is also
    defined by~\eqref{eq:cstone2}. 
  \end{enumerate}
\end{thm}
\begin{rem}
  We could equivalently state the universal
  equation~\eqref{eq:universal} in a somewhat simpler form
  \begin{eqnordisp}
   \frac{\partial}{\partial s} \frac{d}{dt} f(t;s,x,y)= (f*H-H*f),
  \end{eqnordisp} 
  which was already proposed in~\cite{Kisil96a}, but
  it hides the universal nature of $p$-\-mechanical
  bracket~\eqref{eq:u-brackets}.
\end{rem}
\begin{cor}[Consistence of Dynamics]
  \label{co:consistent-dynamics}
  Dynamic defined by  $p$-\-mechanical
  equation~\eqref{eq:universal} and consequently by either its
  derivation---the Heisenberg
  equation~\eqref{eq:Heisenberg}, or the Hamilton
  equation~\eqref{eq:Hamilton}---has the properties
  \begin{enumerate}
  \item The identity $C(0)=A(0)+B(0)$ for three observables will be
    valid through the evolution  $C(t)=A(t)+B(t)$,
    $t\in\Space[+]{R}{}$
  \item It preserve a time independent Hamiltonian.
  \item Corresponding brackets ($\ub{A}{B}$, $\{A,B\}$, $[A,B]$) of
    two observables $A$ and $B$ is again an observable evolving by the
    same equation. 
  \item The identity $C(0)=A(0)B(0)$ for three observables will be
    valid through the evolution  $C(t)=A(t)B(t)$,
    $t\in\Space[+]{R}{}$. 
  \item The Schr\"odinger-Luiville and Hamilton-Heisenberg pictures of 
    motion are equivalent.
  \end{enumerate}
\end{cor}
\begin{proof}
  It is known (see~\cite{CaroSalcedo99}) that the above four
  properties are a direct consequence of those from
  Lemma~\ref{le:ub-Lie}. Again the properties are very well known for
  the quantum commutator and the Poisson brackets.
\end{proof}

Of course, it is not difficult to give a general form of a solution to
the $p$-mechanical equation of motions:
\begin{prop}
  Let 
  \begin{eqnarray}
    f(t;s,x,y) & = & \exp (-t\anti H) f_0(s,x,y) \exp (t\anti H),
    \label{eq:universal-solution}\\
     & = & \exp (-t H_\anti) f_0(s,x,y) \exp (t H_\anti), \nonumber
  \end{eqnarray}
  be a function defined on $\Space{R}{}\times\Space{H}{n}$. Here
  in~\eqref{eq:universal-solution} $H$ is the convolution on
  $\Space{H}{n}$ with a Hamiltonian function $H(s,x,y)$, $\anti$ is
  the anti-derivative operator~\eqref{eq:anti-defn}, and $H_\anti$ is
  the convolution with function $\anti H(s,x,y)$. 

  Then $f(t;s,x,y)$
  from~\eqref{eq:universal-solution} satisfies to the $p$-mechanical
  dynamic equation~\eqref{eq:universal}.
\end{prop}

Note that we never use in the above consideration any kind of limits
and approximations of the type $\myhbar \rightarrow 0$. Both cases of
$\myhbar\neq0$ and $\myhbar=0$ were proven independently without any
references each other. On the other hand this limit does exist in the
induced topology on the dual object $\hHeisen{n}$, i.e. the set of
equivalence classes of unitary irreducible
representation~\cite[\S~7.3]{Kirillov76}) of the Heisenberg
group. This topology was considered for example in~\cite{Kisil96a} and
it was shown that the set of representation $\rho_\myhbar$,
$\myhbar\in (0,\epsilon)$ is dense in the set of representations
$\rho_{(q,p)}$, $p, q\in \Space{R}{n}$. Because we obtain both
equations~\eqref{eq:Heisenberg} and~\eqref{eq:Hamilton} from the same
source~\eqref{eq:universal} we could conclude:
\begin{cor}[The Correspondence Principle]
  Quantum dynamics is dense in classic dynamics, or in 
  loose terms: classic dynamics a limiting case of quantum one.
\end{cor}

\subsection{Example: the Harmonic Oscillator}

\label{sec:exampl-harm-oscill}
We consider ``the lovely pet'' of quantum mechanics---the harmonic
oscillator. Fortunately its consideration within $p$-\-mechanics is 
as well easy.

The well known~\cite[\S~1.6]{MTaylor86} Hamiltonian of a classic
harmonic oscillator is $H_0(q,p)=q^2+p^2$ and in quantum case
Hamiltonian is $H_{\myhbar}=\myhbar(M^2 + D^2)$, where operators $M$
and $D$ defined in~(\ref{eq:M-definition}--\ref{eq:D-definition}). It
easy to find a $p$-\-mechanical Hamiltonian which generates both quantum
and classic ones.
\begin{lem}
  \label{le:ho-equation}
  \begin{enumerate}
  \item Let 
    \begin{eqnordisp}[eq:H-one]
      H(s,x,y)=\delta(s)\delta^{(2)}(x)\delta(y)
      +\delta(s)\delta(x)\delta^{(2)}(y),
    \end{eqnordisp}
    where
    $\delta^{(2)}$ is the second derivative~\cite[\S~III.4.4]{KirGvi82}
    of the Dirac delta function $\delta(x)$.  Then
    $H_\myhbar=\myhbar(M^2 + D^2) $ and
    $H_0(q,p)=q^2+p^2$ are images of $H$ under representations
    $\rho_{\myhbar}$~\eqref{eq:cstone1} and
    $\rho_{(q,p)}$~\eqref{eq:cstone2} correspondingly.
  \item The $p$-\-mechanical equation $\dot{f}=\ub{H}{f}$ of the
        harmonic oscillator is
  \begin{eqnordisp}[eq:p-oscillator]
    \frac{d}{dt} f(t;s,x,y)= 
    2\sum_{j=1}^n \left(x_j \frac{\partial}{\partial y_j} -
      y_j\frac{\partial}{\partial x_j}\right) f(t;s,x,y).
  \end{eqnordisp}
\end{enumerate}
\end{lem}
\begin{proof}
  To establish first statement one verifies images of
  $H(s,x,y)=\delta^{(2)}(x)+\delta^{(2)}(y)$ under representations
  $\rho_{\myhbar}$~\eqref{eq:cstone1} and
  $\rho_{(q,p)}$~\eqref{eq:cstone2} by a direct calculation. We proceed
  with a derivation of the equation~\eqref{eq:p-oscillator}. Let
  \cite[Chap.~1, (1.27)]{MTaylor86}
  \begin{eqnarray}
    X_j^r= \frac{\partial}{\partial x_j}
    +\frac{y_j}{2}\frac{\partial}{\partial s}, &&
    Y_j^r= \frac{\partial}{\partial y_j}
    -\frac{x_j}{2}\frac{\partial}{\partial s}, 
    \label{eq:r-fields}
    \\
    X_j^l= \frac{\partial}{\partial x_j}
    -\frac{y_j}{2}\frac{\partial}{\partial s}, &&
    Y_j^l= \frac{\partial}{\partial y_j}
    +\frac{x_j}{2}\frac{\partial}{\partial s}, \quad
     \textrm{where}\ 1\leq j \leq n,
     \label{eq:l-fields}
  \end{eqnarray} 
  be the left and the right invariant vector fields on
  $\Space{H}{n}$ correspondingly. They generate the right $r(s,x,y)$
  and the left $l(s,x,y)$ shifts on $\FSpace{L}{2}(\Space{H}{n})$
  correspondingly (left invariant vector fields generate right
  shifts and vise verse):
  \begin{eqnarray*}
    \exp\sum_{j=1}^n x_j X_j^r= r(0,x,0), && 
    \exp\sum_{j=1}^n x_j X_j^l= l(0,x,0), 
    \qquad     x=(x_1,\ldots,x_n)\\
    \exp\sum_{j=1}^n y_j Y_j^r= r(0,y,0), &&
    \exp\sum_{j=1}^n y_j Y_j^l= l(0,y,0), 
    \qquad    y=(y_1,\ldots,y_n)
  \end{eqnarray*}
  Then we could express convolutions~\eqref{eq:de-convolution} with
  $\delta^{(2)}$ as second order differential operators:
  \begin{eqnarray*}
    \left(\delta(s)\delta^{(2)}(x)\delta(y)\right)*f
    =\sum_{j=1}^n (X^l_j)^2 f, &&
    \left(\delta(s)\delta(x)\delta^{(2)}(y)\right)*f
    =\sum_{j=1}^n (Y^l_j)^2 f, \\
    f*\left(\delta(s)\delta^{(2)}(x)\delta(y)\right)
    =\sum_{j=1}^n (X^r_j)^2 f, &&
    f*\left(\delta(s)\delta(x)\delta^{(2)}(y)\right)
    =\sum_{j=1}^n (Y^r_j)^2 f.
  \end{eqnarray*}
  Therefore the commutator $[f,H]$ is
  \begin{eqnarray}
    [f,H] & = & f*\left(\delta(s)\delta^{(2)}(x)\delta(y)
      +\delta(s)\delta(x)\delta^{(2)}(y)\right) \nonumber\\
    && {} -
    \left(\delta(s)\delta^{(2)}(x)\delta(y)
      +\delta(s)\delta(x)\delta^{(2)}(y)\right)*f \
    \nonumber\\
     & = &  \sum_{j=1}^n \left( (X^r_j)^2 +(Y^r_j)^2- (X^l_j)^2
       -(Y^l_j)^2 \right) f \nonumber\\
     & = &  \sum_{j=1}^n \left( 
       \left ( X^r_j- X^l_j\right)\left ( X^r_j+ X^l_j\right) +
       \left ( Y^r_j- Y^l_j\right)\left ( Y^r_j+ Y^l_j\right) \right)
     f \nonumber\\
     & = &  \sum_{j=1}^n \left( 
       2y_j\frac{\partial}{\partial s}\frac{\partial}{\partial x_j}
       -2x_j\frac{\partial}{\partial s}\frac{\partial}{\partial y_j}
     \right)
     f \label{eq:ho-transf1} \\
     & = & 2\frac{\partial}{\partial s}  \sum_{j=1}^n \left( 
       y_j\frac{\partial}{\partial x_j}
       -x_j \frac{\partial}{\partial y_j}
     \right)
     f \nonumber
  \end{eqnarray}
  We substitute values from~(\ref{eq:r-fields}--\ref{eq:l-fields}) in
  order to obtain~\eqref{eq:ho-transf1}.
  Finally the $p$-brackets~\eqref{eq:u-brackets} are
  \begin{eqnarray}
    \ub{f}{H} & =& \anti [f,H] \nonumber\\
    & = & \anti\, 2\frac{\partial}{\partial s}  \sum_{j=1}^n \left( 
       y_j\frac{\partial}{\partial x_j}
       -x_j \frac{\partial}{\partial y_j}
     \right)
     f \nonumber\\
    & = & 2\sum_{j=1}^n \left( 
       y_j\frac{\partial}{\partial x_j}
       -x_j \frac{\partial}{\partial y_j}
     \right)
     f  \label{eq:ho-trans2}
  \end{eqnarray}
  Substitution of the last formula~\eqref{eq:ho-trans2} into
  $p$-\-mechanical equation~\eqref{eq:universal}
  proves~\eqref{eq:p-oscillator}. 
\end{proof}

The solution of the equation~\eqref{eq:p-oscillator} is well known.
\begin{lem}
  \label{le:ho-evolution}
  The evolution of an observable $f(t;s,x,y)$ of the $p$-\-mechanical
  harmonic oscillator is given by 
  \begin{eqnarray}
    f(t;s,x,y) & = & f_0(s, x\cos t + y\sin t, -x\sin t + y\cos t)    
    \label{eq:ho-evolution} \\
    & = &  f_0 (s, e^{-it}z), \qquad \textrm{ for }\quad z=x+iy,
    \nonumber
  \end{eqnarray}
  where $f_0(s,x,y)=f(0;s,x,y)$ is the initial value of the observable 
  at $t=0$.
\end{lem}

The above evolution is transparently geometric. In order to preserve
this property in quantum mechanics we introduce in our consideration
the Segal-Bargmann(-Fock) 
space~\cite{Bargmann61,Berezin74,BergCob87,Guillemin84,%
Howe80b,Segal60}.   
Let $\FSpace{L}{2}(\Space{C}{n},d\mu_{n})$ be a space of functions on
\Space{C}{n} which are square-integrable with respect to the Gaussian
measure
\begin{displaymath} 
  d\mu_{n}(z)=\pi^{-n}e^{-z\cdot\overline{z}}dv(z), 
\end{displaymath} 
where $dv(z)=dx\,dy$ is the Euclidean volume measure
on $\Space{C}{n}=\Space{R}{2n}$. The
Segal-Bargmann~\cite{Bargmann61,Segal60} space
$\FSpace{F}{2}(\Space{C}{n})$ is the subspace of
$\FSpace{L}{2}(\Space{C}{n},d\mu_{n})$ consisting of all entire
functions, i.e. functions $f(z)$ that satisfy
\begin{displaymath} 
  \frac{\partial f}{\partial \bar{z}_j}=0, \qquad 1\leq j \leq n. 
\end{displaymath} 

Then the Heisenberg group $\Space{H}{n}$ acts on
$\FSpace{F}{2}(\Space{C}{n})$ by the irreducible unitary representation
\begin{eqnordisp}[eq:fock-represent]
  \beta_{\myhbar}(s,z) f(w)= \exp\left(2is\myhbar + 
    i\sqrt{\myhbar}zw-\modulus{z}^2 \right) 
  f\!\left(w+i\sqrt{\myhbar}\bar{z}\right), 
\end{eqnordisp} 
where $z=x+iy$, $(s,z)\in\Space{H}{n}$. Of course by the
Stone-von Neumann Theorem~\ref{th:Stone-von-Neumann}
representations~\eqref{eq:stone1} and~\eqref{eq:fock-represent} are
unitary equivalent.

\begin{example}
  \label{ex:oscillator-one}
  In the Segal-Bargmann representation~\cite{BergCob87} creation and
  annihilation operators are $a^+_j=z_jI$ and $a^-_j=\partial/\partial
  z_j$, respectively.  The corresponding quantum Hamiltonian of
  harmonic oscillator is obtained by the Bargmann projection
  \begin{equation}\label{eq:euler} 
    T_{H(q,p)}= \frac{1}{2} P_Q \sum_{j=1}^{n}(q_j^2 + 
    p_j^2)I=\frac{1}{2} (nI+\sum_{j=1}^{n}z_j\frac{\partial 
      }{\partial z_j}). 
  \end{equation} 
  The right side of \eqref{eq:euler} is the celebrated Euler operator.
  It generates the well known dynamical group~\cite[Chap.~1,
  (6.35)]{MTaylor86}
  \begin{equation}\label{eq:oscil-dyn} 
    e^{itT_{H(q,p)}} f(z)= e^{int/2} f(e^{it} z), \qquad f(z)\in 
    \FSpace{F}{2}(\Space{C}{n}), 
  \end{equation} 
  which induces rotation of the $\Space{C}{n}$ space. Note that the
  frequency of the above rotation does not depends from $\myhbar$.
  
  The evolution of the classical oscillator is also given by a rotation
  with the same frequency, that of the phase space $\Space{R}{2n}$
  \begin{equation} 
    z(t)=G_t z_0= e^{it}z_0, \qquad z(t)=p(t)+iq(t),\ 
    z_0=p_0+iq_0. 
  \end{equation} 
  The projection $P_Q$ leads to the Segal-Bargmann representation,
  providing a very straightforward correspondence between quantum and
  classical mechanics of oscillators, in contrast to the rather
  complicated case of the Heisenberg representation~\cite[Chap.~1,
  Prop.~7.1]{MTaylor86}.  The powers of $z$ are the eigenfunctions
  $\phi_n(z)=z^n$ of the Hamiltonian~(\ref{eq:euler}), and the integers
  $n$ are the corresponding eigenvalues.
  Either pure or mixed, any initial state of the oscillator remains
  unchanged during the~\eqref{eq:oscil-dyn} evolutions and no
  transitions between states are observed.
\end{example}